\documentclass[conference]{IEEEtran}
\IEEEoverridecommandlockouts

\usepackage{cite}
\usepackage{amsmath,amssymb,amsfonts}
\usepackage{algorithmic}
\usepackage{graphicx}
\usepackage{textcomp}
\usepackage{xcolor, color, soul}
\usepackage[normalem]{ulem}
\usepackage{graphicx}
\usepackage[linesnumbered,lined,boxed,commentsnumbered,ruled]{algorithm2e}
\usepackage{bm}

\def\BibTeX{{\rm B\kern-.05em{\sc i\kern-.025em b}\kern-.08em
    T\kern-.1667em\lower.7ex\hbox{E}\kern-.125emX}}

\begin{document}

\title{Golden Ratio Search: A Low-Power Adversarial Attack for Deep Learning based Modulation Classification}

\author{
\IEEEauthorblockN{Deepsayan Sadhukhan, Nitin Priyadarshini Shankar, Sheetal Kalyani}
\IEEEauthorblockA{\textit{Department of Electrical Engineering} \\
\textit{Indian Institute of Technology Madras, Chennai - 600036, India}\\
email: \{ee20s001@smail, ee20d425@smail, skalyani@ee\}.iitm.ac.in}
}

\maketitle

\begin{abstract}We propose a minimal power white box adversarial attack for Deep Learning based Automatic Modulation Classification (AMC). The proposed attack uses the Golden Ratio Search (GRS) method to find powerful attacks with minimal power. We evaluate the efficacy of the proposed method by comparing it with existing adversarial attack approaches. Additionally, we test the robustness of the proposed attack against various state-of-the-art architectures, including defense mechanisms such as adversarial training, binarization, and ensemble methods. Experimental results demonstrate that the proposed attack is powerful, requires minimal power, and can be generated in less time, significantly challenging the resilience of current AMC methods.
\end{abstract}

\begin{IEEEkeywords}
    Deep Learning, Wireless Communication, Automatic Modulation Classification, Adversarial Attack, and Golden Ratio Search. 
\end{IEEEkeywords}

\section{Introduction}

In recent years, Automatic Modulation Classification (AMC) has been widely used to predict the modulation type of communication signals {in civilian and military applications \cite{meng2018automatic} \cite{iglesias2011automatic}}. The prediction of the modulation scheme based on the received I/Q (In-phase/Quadrature-phase) samples eliminates the need for control information required to communicate the modulation scheme to the Receiver (Rx). In military use cases, identifying a signal's modulation scheme can help detect potential attacks. {Multiple} Deep Neural Networks (DNNs) have been applied {for} AMC and have achieved state-of-the-art (SOTA) performance. RMLResNet \cite{o2018over} is a deep residual network that uses skip connections to allow features to operate at multiple scales and depths through the network. MCNet \cite{huynh2020mcnet} uses specially designed convolutional blocks to learn spatio-temporal correlations. Lightweight \cite{kim2020lightweight} employs bottleneck and asymmetric convolution structures to reduce computational complexity, having the least parameters. RBLResNet \cite{shankar2024binarized} is a binary-quantized neural network that uses rotation to {enable AMC at the edge}.

However, DNNs are vulnerable to adversarial attacks, and their performance drops drastically in their presence. Adversarial attacks are samples crafted by adding a minimal {structured} perturbation to the original signal. The{se minute} perturbations {significantly} lower performance and severely affect security and reliability. Several methods have been proposed to generate adversarial perturbations, with the Fast Gradient Sign Method (FGSM) \cite{DBLP:journals/corr/GoodfellowSS14} {being} a simple yet powerful technique. In \cite{DBLP:conf/iclr/MadryMSTV18}, the authors proposed the Projected Gradient Descent (PGD) attack, a {more} powerful and multi-step version of FGSM. Additionally, \cite{papernot2016limitations} introduced the Jacobian-based saliency map attack (JSMA), utilizing forward derivatives to create adversarial perturbations. Carlini and Wagner \cite{carlini2017towards} further expanded the arsenal of attack methods by exploring three distinct distance metrics {($L_1$, $L_2$, and $L_\infty$)} for generating adversarial perturbations. Among different attacks, minimum power adversarial attacks are crucial because they maximize the effectiveness of adversarial examples while minimizing the changes made to the input data \cite{ke2023minimum}.

Recently, several works have {studied the impact of} adversarial attacks in AMC. The authors of \cite{lin2020adversarial} explore the performance of attack methods on AMC, measure the effectiveness of adversarial attacks, and provide the empirical evaluation of the reliabilities of CNNs. In \cite{sadeghi2018adversarial}, the authors propose practical methods for crafting white-box and universal black-box adversarial attacks. Using bisection search, they demonstrated that these attacks can significantly degrade classification performance with minimal input perturbations. Their findings highlight that these adversarial attacks are far more effective than traditional jamming attacks. {In \cite{kim2020over}, the authors consider the effects of the wireless channel between the adversary and the receiver and use the algorithm in \cite{sadeghi2018adversarial} to maximize perturbation at the receiver. Further, in \cite{kim2021channel}, the authors generalize the algorithm in \cite{sadeghi2018adversarial} to a broadcast attack where the channel information is considered across multiple receivers.}

The Golden Ratio (GR) was first applied in machine learning in \cite{jaeger2021golden} to find the optimal scaling factor to minimize an information-theoretical loss function based on cross-entropy. {The GR was used in \cite{jaeger2021golden} to calculate theoretical values for the regularization parameters, namely, learning rate and momentum weights, resulting in efficient model training.} In this work, we {design} adversarial attacks that consume low power without compromising the attack performance. We construct an attack mechanism inspired by targeted FGSM {and incorporate} a Golden Ratio Search (GRS) {method} to find the least perturbation strength that fools the model. The main contributions of the work are as follows: 

\begin{enumerate}
    \item We propose a novel attack method that uses GRS optimization to generate a minimum power adversarial attack for AMC. 
    \item We evaluate the performance of SOTA architectures specifically crafted for AMC. 
    \item We compare the proposed method with the existing attack methods and evaluate its efficiency against various defense mechanisms.
\end{enumerate}



\section{System Model}


\label{sec:systemmodel}

The transmitted signal $\mathbf{s}$ has a constant normalized symbol rate across all modulation schemes. The received signal $\mathbf{x}$ is given by 
\begin{equation}
    \mathbf{x}=\mathbf{s} \ast \mathbf{h} + \mathbf{n}    
\end{equation}
where $\mathbf{h}$ is the channel vector, $*$ denotes the convolution operator, and the noise vector $\mathbf{n}$ added to the received signal is complex AWGN, with each sample distributed as $\mathcal{CN}(0, N_0)$. We consider the following channel effects - Sample Rate Offset (SRO), Center Frequency Offset (CFO), selective fading models (Rician and Rayleigh), and finally, the Additive White Gaussian Noise (AWGN). The dataset models all these channel effects. In this work, we consider the challenging RML$2018.01$a (R: Radio; ML: Machine Learning) dataset \cite{o2018over} available at deepsig\footnote{https://www.deepsig.ai/datasets}. 

The dataset comprises both real-world measurements and simulated channel effects. The synthetic data was produced using software-defined radio programmed with GNU radio \cite{blossom2004gnu}. The dataset has $24$ modulations schemes, namely OOK, 4ASK, 8ASK, BPSK, QPSK, 8PSK, 16PSK, 32PSK, 16APSK, 32APSK, 64APSK, 128APSK, 16QAM, 32QAM, 64QAM, 128QAM, 256QAM, AM-SSB-WC, AM-SSB-SC, AM-DSB-WC, AM-DSB-SC, FM, GMSK, OQPSK, with $26$ evenly spaced bins in Signal-to-Noise Ratio (SNR), ranging from $-20$ to $30$dB. The set comprises $2,555,904$ I/Q (in-phase/quadrature) signals, each of length $1024$ (array shape is $2 \times 1024$). The In-phase (I) and Quadrature (Q) components of the signal represent all signals as two-dimensional reals even though they are all complex. 

Besides channel and noise effects, adversarial attacks can further degrade the transmitted signal. In an adversarial setting, the attacker adds an adversarial component to the clean signal denoted by $\mathbf{r_x}$, and the received signal is 
\begin{equation}
    \mathbf{x_{adv}}=\mathbf{x}+\mathbf{r_x}
\end{equation}
Based on the availability of model information (weights and gradients) to the attackers, there are two types of attacks: (i) white box and (ii) black box. In black box attacks, the adversary can only access the outputs from the model. {White-box attacks are more powerful than black-box attacks; hence, we propose minimal-power white-box adversarial attacks in this work.}

\begin{algorithm}
\SetKwInOut{Input}{Inputs}\SetKwInOut{Output}{Output}
\SetKw{KwTo}{in}
\caption{Crafting an adversarial example}\label{alg:two}
\Input{\begin{itemize}
    \item input  $\mathbf{x} \text { and its label } l_{\text {true }}$
    \item the model $f(., \boldsymbol{\theta})$
    \item minimum tolerance $tol$
    \item maximum allowed perturbation norm  $p_{\max }$
    \end{itemize}}
\Output{adversarial perturbation of the input, i.e., $\mathbf{r_x}$}

$\text {Initialize: } \bm{\varepsilon} \leftarrow \mathbf{0}^{C \times 1},  \phi \leftarrow \frac{ \sqrt{5}-1}{2} $

\For{$ \text{class-index}$  \KwTo $\text{range}(C)$}{

$\varepsilon_{\max } \leftarrow p_{\max }, \varepsilon_{\min } \leftarrow 0$


\hbox{$\mathbf{r}_{\text{norm}}\hspace{-0.07cm}=\hspace{-0.07cm}\left(\left\|\nabla_{\mathbf{x}} L\left(\mathbf{x}, \mathbf{e}_{\text{\textit{class-index}\hspace{-0.07cm} }}\right)\right\|_2\right)^{-1} \nabla_{\mathbf{x}} L\left(\mathbf{x}, \mathbf{e}_{\text{\textit{class-index}\hspace{-0.02cm}}}\right)$}

\While{$\varepsilon_{\max }-\varepsilon_{\min }>tol$}{
$\varepsilon_{\text {ave }} \leftarrow \varepsilon_{\min }+\left (\varepsilon_{\max }-\varepsilon_{\min }\right) * \phi$

$\mathbf{x}_{\text {adv }} \leftarrow \mathbf{x}-\varepsilon_{\text {ave }} \mathbf{r}_{\text {norm }}$

  \eIf{$\hat{l}\left(\mathbf{x}_{a d v}\right)==l_{\text {true }}$}{
  $\varepsilon_{\min } \leftarrow \varepsilon_{\text {ave }}$

  }{$\varepsilon_{\max } \leftarrow \varepsilon_{\text {ave }}$
  }
  }
$[\bm{\varepsilon}]_{\text {\textit{class-index} }}=\varepsilon_{\text {ave }}$
}
$ \text{\textit{target-class}}={\arg \min } \ \bm{\varepsilon} \text { and } \varepsilon^*=\min  \bm{\varepsilon}$

$\mathbf{r}_{\mathbf{x}}=-\frac{\varepsilon^{\mathbf{*}}}{\left\|\nabla_{\mathbf{x}} L\left(\mathbf{x}, \mathbf{e}_{\text {\textit{target-class} }}\right)\right\|_2} \nabla_{\mathbf{x}} L\left(\mathbf{x}, \mathbf{e}_{\text {\textit{target-class} }}\right)$
\end{algorithm}\label{Algo}
\section{Proposed Method}

The objective of this proposed method is to leverage the GRS algorithm as an optimization technique for crafting minimal power adversarial attacks in the context of AMC. It helps efficiently identify the optimal perturbations in the input space that lead to misclassifications while minimizing the impact on power consumption. 


{The steps of the proposed method are given in Algorithm 1. The algorithm generates an adversarial sample by crafting a perturbation for a given input to fool the classifier at the receiver. It takes an input sample $\mathbf{x}$, its true label $l_{\text{true}}$, total number of class labels $C$, the model $f(., \boldsymbol{\theta})$, a tolerance $tol$, and a maximum perturbation norm $p_{\max}$. The algorithm iterates over each class, calculating the gradient of the loss function $L\left(\mathbf{x}, \mathbf{e}_{\text{\textit{class-index}\hspace{-0.02cm}}}\right)$ with respect to the input $\mathbf{x}$, where $\mathbf{e}_{\text{\textit{class-index}}}$ is the one-hot encoded class. It then searches for the smallest perturbation $\varepsilon_{\text{ave}}$ within the allowable range that still causes the model to misclassify $\mathbf{x}$ when perturbed. The algorithm uses the golden ratio $\phi=\frac{\sqrt{5}-1}{2}$ to narrow down this range efficiently by splitting the search space in the ratio $\phi:1$. Finally, it selects the target class that minimizes this perturbation and generates the adversarial perturbation $\mathbf{r_x}$ based on the gradient, which causes the model to predict the wrong label when added to the input. A smaller $\varepsilon^{\mathbf{*}}$ indicates a weaker perturbation. Thus, as in \cite{sadeghi2018adversarial}, the adversary can create targeted attacks for all $C-1$ modulation types and chooses the target modulation that uses the least power.}
Therefore, it provides fine-grained adversarial perturbations while relying on the computationally efficient FGSM as the algorithm's core. It converges faster than \cite{sadeghi2018adversarial} due to the GRS optimization method.

We will validate the proposed method's efficacy and robustness through experiments, including evaluating adversarial attacks on baseline and adversarially enhanced models. To evaluate the robustness of a classifier $f$  to adversarial perturbations, we use the average robustness \cite{moosavi2016deepfool}, denoted by $\hat{\rho}_{\text{adv}}\left(f\right)$. This metric quantifies how resistant the classifier is to small perturbations designed to fool it. The average robustness is computed as 
\begin{equation}
    \hat{\rho}_{\text{adv}}(f) = \frac{1}{|D|} \sum_{x \in D} \frac{\| \hat{r}(x) \|_\infty}{\| x \|_\infty}
\end{equation}

where, \( |D| \) is the size of the test set \( D \), \( x \) represents an individual data point in the test set \( D \), \( \hat{r}(x) \) is the estimated minimal perturbation required to misclassify \( x \), which is obtained using the proposed method, and \( \| \cdot \|_\infty \) denotes the $L_\infty$ norm. In simpler terms, for each data point \( x \) in the test set, we calculate the ratio of the $L_\infty$ norm of the minimal perturbation \( \hat{r}(x) \) to the $L_\infty$ norm of the original data point \( x \). The ratio \( \frac{\| \hat{r}(x) \|_\infty}{\| x \|_\infty} \) effectively quantifies the smallest perturbation relative to the input's magnitude, sufficient to change the classifier's decision. The average robustness metric, \( \hat{\rho}_{\text{adv}}(f) \), is calculated by taking an average of these ratios over all data points in the test set. It can be interpreted as an attacker's minimum power required to fool the classifier. This interpretation arises from the nature of the $L_\infty$ norm, which measures the maximum change to a single feature in the input data.  Thus, a higher value of \( \hat{\rho}_{\text{adv}}(f) \) indicates that the classifier requires larger perturbations (more power) to be fooled, demonstrating greater robustness. Conversely, a lower value of \( \hat{\rho}_{\text{adv}}(f) \) suggests that the classifier can be fooled with smaller perturbations (low power), indicating lower robustness.

\section{Experiments}

\begin{table*}[ht]
\caption{A Comparison of Accuracy, Robustness, and Attack Time across various SOTA methods facing adversarial attacks at $10$dB SNR on RML2018.01a dataset.\label{tab:allmethods}}
  \begin{center}
    \begin{tabular}{|p{2.7cm}|p{0.8cm} |p{0.8cm}| p{0.8cm}| p{0.8cm}|p{0.8cm}|p{0.8cm}|p{0.8cm}|p{0.8cm}|p{0.8cm}|p{0.8cm}|p{0.8cm}|p{0.8cm}|}
 \hline
{\textbf{Model}} & \multicolumn{4}{|c|}{\textbf{Accuracy (\%)}} & \multicolumn{4}{|c|}{\textbf{Robustness}} & \multicolumn{4}{|c|}{\textbf{Attack Time (s)}} \\
\cline{2-13}
& GRS &FGSM & PGD& CW & GRS &FGSM & PGD& CW & GRS &FGSM & PGD& CW \\
\hline
Lightweight                             & \textbf{23.434} & 32.245 & 30.417          & 29.945           & \textbf{0.0300} & 0.0358 & 0.0358             & 0.3102           & 6.966  & 0.036 & 0.294 & 14.871 \\
MCNet                                   & \textbf{13.038} & 17.874 & 16.228          & 14.745           & \textbf{0.0346} & 0.0358 & 0.0358             & 0.3177           & 11.001 & 0.045 & 0.518  & 23.090 \\
LResNet                                 & \textbf{24.072} & 33.168 & 31.616          & 32.985           & \textbf{0.0339} & 0.0358 & 0.0358             & 0.3105           & 7.392  & 0.035 & 0.362  & 16.926 \\
RBLResNet                               & \textbf{47.535} & 49.941 & 61.719          & 70.23            & \textbf{0.0357} & 0.0358 & 0.0358             & 0.3060           & 13.162 & 0.073  & 0.847 & 34.344 \\
RMLResNet                               & 27.982          & 45.405 & 17.074 & \textbf{6.645}   & 0.0368          & \textbf{0.0358} & \textbf{0.0358}   & {0.3154}  & 8.642  & 0.039 & 0.409 & 19.261 \\
MCNet (AT$_{\text{FGSM}}$)              & \textbf{36.013} & 36.732 & 49.084          & 40.715           & \textbf{0.0355} & 0.0358 & 0.0358             & 0.3144           & 7.615  & 0.045 & 0.519 & 22.694 \\
MCNet (AT$_{\text{PGD}}$)               & \textbf{27.669} & 31.006 & 37.716          & 32.416           & \textbf{0.0326} & 0.0358 & 0.0358             & 0.3188           & 10.244 & 0.045 & 0.519 & 23.017  \\
RMLResNet (AT$_{\text{FGSM}}$)          & 45.848          & 51.571 & 51.449          & \textbf{29.043}           & 0.0370          & \textbf{0.0358} & \textbf{0.0358}             & 0.3176           & 6.050  & 0.037  & 0.402 & 18.702 \\
RMLResNet (AT$_{\text{PGD}}$)           & 36.298          & 37.748 & 44.231          & \textbf{26.291}          & \textbf{0.0339}         & 0.0358 & 0.0358             & 0.3248           & 8.350  & 0.037 & 0.400 & 20.017 \\
LResNet (AT$_{\text{FGSM}}$)            & \textbf{30.986} & 33.603 & 36.854          & 36.371           & \textbf{0.0324} & 0.0358 & 0.0358             & 0.3094           & 7.095   & 0.040 & 0.383 & 16.571 \\
RBLResNet-Bag2                          & \textbf{47.661} & 53.668 & 59.963          & 76.789           & \textbf{0.0357} & 0.0358 & 0.0358             & 0.3061           & 12.701 & 0.077 & 1.733  & 75.433 \\
\hline
\end{tabular}
\end{center}
\end{table*}

 \begin{table*}[ht]
 \caption{A Comparison of Robustness across SNR. For FGSM/PGD, a fixed perturbation strength $\left(\epsilon=0.05\right)$ was used.\label{tab:ROBvSNR}}
  \begin{center}
    \begin{tabular}{|p{3.215cm}|p{0.85cm} |p{0.85cm}| p{0.85cm}| p{0.85cm}|p{0.85cm}|p{0.85cm}|p{0.85cm}|p{0.85cm}|p{0.85cm}|p{0.85cm}|p{0.85cm}|}
 \hline
{\textbf{Attack (Model)}} & \multicolumn{11}{|c|}{\textbf{SNR}}  \\
\cline{2-12}
& $0$ dB & $2$ dB & $4$ dB& $6$ dB & $8$ dB &$10$ dB & $12$ dB& $14$ dB & $16$ dB &$18$ dB & $20$ dB \\
\hline
GRS (LResNet) & \textbf{0.0152} &\textbf{0.0183}&\textbf{0.0206}&\textbf{0.0229}&\textbf{0.0255}&\textbf{0.0339}&\textbf{0.0318}&\textbf{0.0325}&\textbf{0.0334}&\textbf{0.0330}&\textbf{0.0336}\\
GRS (RBLResNet) & 0.0238 & 0.0263 &0.0293 &0.0320&0.0334&0.0356&0.0354&0.0355&0.0350&0.0357&0.0360\\
FGSM/PGD & 0.0250 & 0.0271& 0.0296 & 0.0323 & 0.0340&0.0357& 0.0359&0.0362&0.0364&0.0363&0.0365 \\
\hline
\end{tabular} 
\end{center}

\end{table*}

 \begin{table}[ht]
 \caption{A Comparison of \cite{sadeghi2018adversarial} and GRS accross models.\label{tab:BSSvGSS}}
  \begin{center}
    \begin{tabular}{|p{1.8cm}|p{0.9cm} |p{1.4cm}| p{0.95cm}| p{1.4cm}|}
 \hline
{\textbf{Model}} & \multicolumn{2}{|c|}{\textbf{Accuracy (\%)}}  &  \multicolumn{2}{|c|}{\textbf{Attack Time (s)}}\\
\cline{2-5}
&\cite{sadeghi2018adversarial}&GRS \newline(proposed) &\cite{sadeghi2018adversarial}&GRS \newline(proposed) \\
\hline
Lightweight	&23.639&	\textbf{23.434} & 8.369&	\textbf{6.966} 	\\
MCNet & 13.236 & \textbf{13.038} &  13.237& \textbf{11.001} \\ 
LResNet & 24.072 & \textbf{24.012} &  9.369& \textbf{7.392}\\
RBLResNet & 47.535 &  \textbf{46.936} & 14.937& \textbf{13.162}\\ 
RMLResNet & 28.124  & \textbf{27.982} & 9.867& \textbf{8.642} \\
\hline
\end{tabular}
\end{center}

\end{table}

\subsection{Experiment Setup}

All the experiments were performed on a system with a $2.90$ GHz CPU, $128$ GB RAM, and an NVIDIA A100 GPU. {All the attacks used for comparison are modeled using the Cleverhans Attack Tool kit \cite{papernot2016technical}.}  For FGSM-based attacks, {we have used perturbation strength/step-size $\left(\epsilon\right)$ of $0.05$ similar to \cite{shankar2024binarized}.} In the case of PGD for a given perturbation strength, we have used a step-size of $0.05$ and the number of iterations $\left(\eta\right)$ as $10$. We used $p_{max}=0.05$ for the GRS attack with $C=24$ for the RML2018.01A dataset. {Adversarial samples for the Carlini-Wagner (CW) attack are {created} with $100$ iterations and an initial constant {$\left(\delta_i\right)$} of $10^{-3}$\cite{carlini2017towards}}. All models are trained for $200$ epochs. In the dataset, $75\%$ of the samples are used for training and the rest for testing. We use the categorical cross-entropy as the loss function. In the case of adversarially trained models, $50\%$ of samples taken for training were generated from the adversarial attack method. 



\subsection{Experiment Results}
\begin{figure}
    \centering
    \includegraphics[scale=0.67]{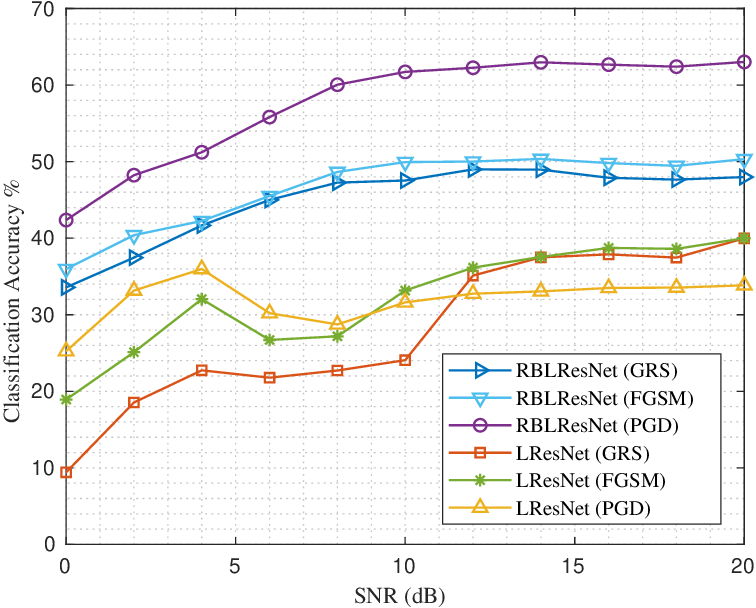}
    \caption{Accuracy comparison in the presence of Adversarial attacks across SNRs.\label{fig:AccvSNR}} 
\end{figure}
On RML2018.01a, the proposed attack method is compared with different attacks in ML literature on several SOTA architectures for AMC in Table \ref{tab:allmethods}. We {consider} performance parameters {such as} adversarial accuracy, robustness, and attack time. The attack time is calculated as {the time} required to generate adversarial samples per batch. GRS outperforms attack methods for architectures with real-valued weights like Lightweight, MCNet, and LResNet. Despite having the highest clean accuracy among the real-valued architectures, LResNet \cite{shankar2024binarized} shows significantly lower adversarial accuracy ($24.072\%$) and robustness ($0.0339$) under the GRS attack {when compared with the accuracies for} FGSM, PGD, and CW attacks. This trend is consistent across other real-valued architectures in Table \ref{tab:allmethods}. {In all the tables, the numbers in bold indicate the most effective attack.}

  
 Even binarized networks {such as} RBLResNet that are inherently robust against adversarial attacks \cite{jia2020efficient}, {have} the lowest adversarial accuracy ($47.535\%$) against GRS. RBLResNet also has the lowest robustness ($0.0357$) for GRS, as seen in Table \ref{tab:allmethods}. {RBLResNet-Bag2, an ensemble-based architecture introduced in \cite{shankar2024binarized}, is designed to enhance weaker networks through the use of bagging. This approach involves Lipschitz bagging, as outlined in \cite{tholeti2022robustwaystackbag}, where the authors leveraged the connection between the local Lipschitz constant and adversarial robustness.}
 The proposed GRS attack method is tested on RBLResNet-Bag2, resulting in lower adversarial accuracy ($47.661\%$)  and robustness ($0.0357$) {when compared with} FGSM, PGD, and CW attacks. 
 
 We {then use} adversarially trained architectures \cite{DBLP:conf/iclr/KurakinGB17} to compare further and test our attack method. MCNet, when adversarially trained with FGSM and PGD samples, gives an adversarial accuracy of $36.013\%$ and $27.669\%$ for GRS attack, which is {again a} lower {accuracy}. Even LResNet, when adversarially trained with FGSM samples, gives the lowest adversarial accuracy for GRS. RMLResNet, when adversarially trained with FGSM and PGD samples, gives an adversarial accuracy of $45.858\%$ and $36.298\%$, respectively, against GRS attacks. The proposed attack method {is} a more potent adversary than FGSM and PGD, {however,} at the expense of the higher time required to craft the adversarial samples. {However, in comparison to both CW and \cite{sadeghi2018adversarial}, the time required to craft the attack is lower, and the attack is more potent as evinced by Tables \ref{tab:allmethods} and \ref{tab:BSSvGSS}. }
 
Additionally, {Fig.} \ref{fig:AccvSNR} shows the attack efficiency across SNRs compared to FGSM/PGD attacks for RBLResNet and LResNet. We see the models have a lower classification accuracy against the proposed method at lower SNRs. The corresponding robustness values, which measure perturbation strength, are tabulated in Table \ref{tab:ROBvSNR}. The table shows that the GRS attack on LResNet has the lowest robustness across a range of SNRs. Further, in  Table \ref{tab:BSSvGSS}, we compare the proposed method with \cite{sadeghi2018adversarial} and find that the proposed method outperforms \cite{sadeghi2018adversarial} both in terms of accuracy and attack time at $10$dB SNR. 

\section{Conclusion}

The proposed method introduces a novel adversarial attack method for deep learning models in AMC using the GRS technique. Through comprehensive experiments, we demonstrated that the GRS method effectively balances adversarial accuracy, sample generation time, and perturbation strength, outperforming established attacks like FGSM, PGD, and CW. Additionally, GRS achieves faster convergence than traditional binary search optimization \cite{sadeghi2018adversarial} while maintaining a high attack success rate. Despite defensive strategies such as binarization and adversarial training, the GRS method remains effective, emphasizing the need for ongoing advancements in defense mechanisms.

\bibliographystyle{IEEEtran}
\bibliography{refs_Gsch}


\end{document}